\documentclass{report}
\usepackage{../style/infrastructure}
\usepackage{latex2oo}
\def\Image#1#2#3{\par \BB{Image: #1 - #2} \par}
\begin{document}
  \nonumchapter{The \DaVinci\ Series}
\Version{30-08-05}

The subtitle of the \DaVinci\ series of lecture notes is \emph{\DaVinciSub}.
On the one hand, this series of lecture notes takes a fundamental view
(\emph{craft}) on the field information systems engineering. At the same time,
it does so with an open eye to practical experiences (the \emph{art}) gained
from information system engineering in industry.

The kinds of information systems we are interested in range from personal
information appliances to enterprise-wide information processing. Even more, we
regard an information system as a system that ``handles'' information, where
``handling'' should be interpreted in a broad fashion. The actors that do this
``handling'' can be computers, but can equally well be other ``symbol wielding
machines'', but can also be humans. The mix of humans and computers/machines in
information systems makes the field of information system engineering
particularly challenging.

The concept of ``information'' itself is very much related to the concepts of
\emph{data}, \emph{knowledge} and \emph{communication}. Based
on~\cite{1998-Lindgreen-FRISCO}, we will (throughout the \DaVinci\ series) use
the following definitions: 
\begin{Definitions}
   Any \DefinitionName{representation}{ls} in some language. 
\DefinitionName{data}{us} is therefore simply a collection of symbols 
that may, or may not, have some meaning to some 
\DefinitionName{actor}{ls}.

   The \DefinitionName{knowledge}{ls} increment brought about when a 
\DefinitionName{human-actor}{ls} receives a 
\DefinitionName{message}{ls}. In other words, it is the difference 
between the \DefinitionName{conception}{lp} held by a 
\DefinitionName{human-actor}{ls} \emph{after} interpreting a received 
\DefinitionName{message}{ls} and the \DefinitionName{conception}{lp} 
held beforehand.

   A relatively stable, and usually mostly consistent, set of 
\DefinitionName{conception}{lp} posessed by a single (possibly 
composed) \DefinitionName{actor}{ls}. In more popular terms: ``an 
\DefinitionName{actor}{ls}'s picture of the world''.

   An exchange of \DefinitionName{message}{lp}, i.e.\ a sequence of mutual 
and alternating \DefinitionName{message}{ls} transfers between at least 
two \DefinitionName{human-actor}{lp}, called 
\DefinitionName{communication}{ls} partners, whereby these 
\DefinitionName{message}{lp} represent some 
\DefinitionName{knowledge}{ls} and are expressed in languages 
understood by all \DefinitionName{communication}{ls} partners, and 
whereby some amount of \DefinitionName{knowledge}{ls} about the 
\DefinitionName{domain}{ls} of \DefinitionName{communication}{ls} and 
about the action context and the \DefinitionName{goal}{ls} of the 
\DefinitionName{communication}{ls} is made present in all 
\DefinitionName{communication}{ls} partners.

\end{Definitions}
When referring to an information system, we therefore really refer to systems
that enable the communication/sharing of knowledge by means of the
representation (by \PDN{human-actor}), storage, processing, retrieval, and
presentation (to \PDN{human-actor}) of the underlying representations (data).
This also implies that we will treat \emph{information retrieval systems},
\emph{knowledge-based systems}, \emph{groupware systems}, etc., as special
classes of information systems.

The lecture notes in the \DaVinci\ series have been organized around four key
key processes in an information system's life-cycle:
\begin{description}
   A process leading to a \DefinitionName{definition-description}{ls}. 
\Where{definition}

   A process leading to a \DefinitionName{design-description}{ls}.
 
 \Where{design}

   The combination of a \DefinitionName{construction-process}{ls} and a 
\DefinitionName{deployment-process}{ls}. \Where{construction-process} 
\Where{deployment-process}

   The processes which tie \DefinitionName{definition}{ls}, 
\DefinitionName{design}{ls} and \DefinitionName{deployment}{ls} and to 
the explicit and implicit needs, desires and 
\DefinitionName{requirement}{lp} of the usage context. Issues such as: 
business/IT \DefinitionName{alignment}{ls}, 
\DefinitionName{stakeholder}{lp}, limiting \DefinitionName{design}{ls} 
freedom, negotiation between \DefinitionName{stakeholder}{lp}, 
enterprise \DefinitionName{architecture}{lp}, 
\DefinitionName{stakeholder}{ls} \DefinitionName{communication}{ls}, 
and outsourcing, typify these processes.

   \DefinitionName{modeling}{us} of the \DefinitionName{domain}{lp} that 
are relevant to the \DefinitionName{information-system}{ls} being 
developed. The resulting \DefinitionName{model}{lp} will typically 
correspond to \emph{ontologies} of the \DefinitionName{domain}{lp}. 
These \DefinitionName{domain}{lp} can pertain to the 
\DefinitionName{information}{ls} that will be processed by the 
\DefinitionName{information-system}{ls}, the processes in which the 
\DefinitionName{information-system}{ls} will play a role, the 
processing as it will occur inside the 
\DefinitionName{information-system}{ls}, etc. Understanding (and 
\DefinitionName{modeling}{ls}) these \DefinitionName{domain}{lp} is 
fundamental to the other activities in 
\DefinitionName{information-system-engineering}{ls}.

\end{description}
For each these aspects, attention will be paid to relevant theories, methods
and techniques to execute the tasks involved. When put together, these aspects
can be related as depicted in figure~\ref{davinci}.  Note that we regard
maintenance of systems as being functionality that should be designed ``into''
the system. If a system needs to be maintained, and in most cases one indeed
wants to, then the maintenance should be designed into the workings of this
system and/or its context.
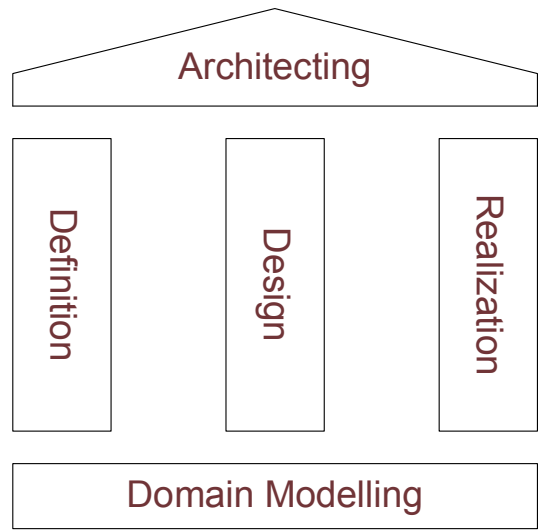
\Image{7cm}{Aspects of Information Systems Engineering}{davinci}

The use of the name \DaVinci\ originates from earlier
work~\cite{1998-Proper-DaVinci} done on architecture-driven information systems
engineering. The work reported in~\cite{1998-Proper-DaVinci} was the result of
a confrontation between industrial practice and a theoretical perspective on
information systems and their evolution~\cite{1994-Proper-EvolvConcModels}. The
result was a shared vision on the architecture-driven development of
information systems by a Dutch IT consultancy firm. In this shared vision, a
foundation was laid for an integrated view on information system engineering.
At that stage, the name ``\DaVinci'' was also selected. Not as some artificial
acronym, but rather to honor an inspiring artist, scientist, inventor and
architect. To us he personifies a balance between art and engineering, between
human and technology.

After the development of the first \DaVinci\ version, a more elaborate
version~\cite{2004-Proper-DaVinci} was developed at the Radboud University
Nijmegen in the form of lecture notes associated to a course on
\emph{Architecture \& Alignment}. In this version, a more fundamental outlook
on information system development was added to complement the practical
orientation of the first version.

As a third step, we have now taken on the underlying philosophy of the first
two \DaVinci\ documents, and used this as the source of inspiration to shape an
entire line of lecture notes for a number of mutually related courses on
different aspects of information systems engineering. In making this step we
have also been able to anchor some of the fundamental research results from the
co-authors, on subjects such as information modeling~\cite{1991-Bommel-PredMod,
1991-Bommel-CSTransform, 1993-Hofstede-PSM, 1993-Hofstede-LISA-D,
1994-Proper-EvolvPSM, 1995-Brouwer-ORMKernel, 1995-Campbell-Abstraction,
1996-Proper-CDMKernel, 1995-Proper-EvolvAM, 1996-Frederiks-Logbooks,
1997-Hofstede-Grammar, 1997-Proper-Versions, 1997-Hoppenbrouwers-NLStructures,
2002-Frederiks-Infogrammars, 2004-Frederiks-InformationModeling}, information
retrieval~\cite{1990-Bruza-AssessingQuality, 1991-Bruza-Retlogic,
1991-Bruza-StratHypmed, 1996-Berger-SemRel, 1998-Wondergem-IndExprMatching,
1999-Sarbo-MeaningExtraction, 1999-Proper-Infons, 2001-Papazoglou-DigiLib,
2001-Wondergem-LogicLinguistic}, (enterprise) information
architecture~\cite{2004-Jonkers-ArchiConcepts} and information system
engineering~\cite{1999-Proper-ISPL-LSM, 2004-VeldhuijzenVanZanten-Rational}
into the core of the \DaVinci\ series.

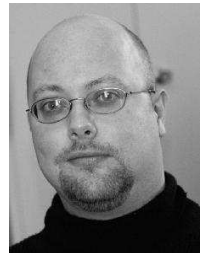
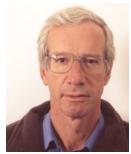
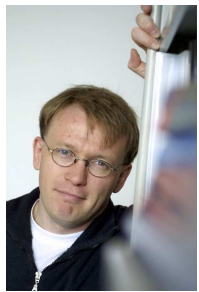
\end{document}